\documentclass[epj]{svjour}
\usepackage{graphics}
\usepackage{hyperref}
\usepackage{color}
\usepackage{amssymb,bm,calc}

\begin{document}

\title{Strange hadron production in pp, pPb and PbPb collisions at LHC energies}


\author{Kapil Saraswat\inst{1}
\and Prashant Shukla\inst{2,3}$^{\ast}$
\and Vineet Kumar \inst{2}
\and Venktesh Singh\inst{1}.}
              
\institute{Department of Physics, Institute of Science, 
Banaras Hindu University, Varanasi 221005, India.
\and 
Nuclear Physics Division, Bhabha Atomic Research Center, Mumbai 400085, India.
\and
Homi Bhabha National Institute, Anushakti Nagar, Mumbai 400094, India.}

\date{Received: date / Revised version: date}

\abstract{
  We present a systematic analysis of transverse momentum $(p_{T})$ spectra 
of the strange hadrons in different multiplicity events produced in pp collision 
at $\sqrt{s}$ = 7 TeV, pPb collision at $\sqrt{s_{NN}}$ = 5.02 TeV and PbPb collision 
at $\sqrt{s_{NN}}$ = 2.76 TeV. Both the single and differential freeze out scenarios 
of strange hadrons $K^0_s$, $\Lambda$ and $\Xi^-$ are considered while fitting using 
a Tsallis distribution which is modified to include transverse flow. The $p_{T}$ 
distributions of these hadrons in different systems are characterized in terms of 
the parameters namely, Tsallis temperature $(T)$, power $(n)$ and average transverse 
flow velocity $(\beta)$. It is found that for all the systems, transverse flow 
increases as we move from lower to higher multiplicity events. 
In the case of the differential freeze-out scenario, the degree of thermalization 
remains similar for events of different multiplicity classes in all the three systems.
The Tsallis temperature increases with the mass of the hadrons and also increases 
with the event multiplicity in pp and pPb system but shows little variation with 
the multiplicity in PbPb system.  
In the case of the single freeze-out scenario, the difference between small systems 
(pp, pPb) and PbPb system becomes more evident. 
The high multiplicity PbPb events show higher degree of thermalization as compared 
to the events of pp and pPb systems. The trend of variation of the temperature in 
PbPb system with event multiplicity is opposite to what is found in the pp and 
pPb systems.
} 
\maketitle


\footnote{$^{\ast}$ E-mail: pshuklabarc@gmail.com}

\section{Introduction}
   The heavy ion collisions at RHIC and LHC aim to create matter with high energy 
density required for the formation of Quark Gluon Plasma (QGP) ~\cite{Bjorken:1982qr} 
\cite{Ullrich:2013qwa} \cite{Gyulassy:2004zy}. 
  The quark gluon matter once formed, presumably with local thermal equilibrium 
expands, cools and undergoes a phase transition to hadronic matter. The hadronic 
matter continues to expand and once the mean free path of hadrons becomes bigger 
than the system size, they decouple from the system and move to the detectors. 
 The transverse momentum ($p_T$) spectra of hadrons reflect the condition of the 
system at the time of freeze-out. 
The $p_T$ spectra of hadrons such as pions, kaons, protons and strange 
baryons have been a very useful tool to study particle production mechanisms, 
thermalization and collective effects in a large system~
\cite{Hirano:2003pw} \cite{Fries:2008hs}. 
At LHC, three different types of collisions: proton-proton (pp), proton-lead (pPb), 
and lead-lead (PbPb) are performed at different center of mass energies.

  The nucleus at high energy is considered to be a system of very dense partons. 
These partons can undergo simultaneous and independent scatterings during initial 
collisions and produce large number of particles in the final state. The 
thermalization time scale commonly assumed for the QGP created in nuclear 
collisions is significantly less than 1 fm/$c$~\cite{Muller:2013dea} which is 
small compared 
to the size of a heavy nucleus and thus it is expected that the system thermalizes. 
There have been discussions of collectivity ~\cite{Li:2012hc} in pp collisions also,  
after the observation of long range rapidity correlations in high multiplicity pp 
events at LHC energy~\cite{Khachatryan:2010gv}. 
A Boltzmann-Gibbs Blast Wave model fit has been performed on strange and non-strange 
hadrons produced in pp collisions in the Ref.~\cite{Ghosh:2014eqa} concluding that 
there is a collective transverse flow in small systems.

 The Tsallis distribution~\cite{Tsallis:1987eu} \cite{Biro:2008hz} describes 
near thermal systems in terms of two parameters, temperature $T$ and parameter 
$q$ which measures temperature fluctuation or degree of non-thermalization. 
It is known \cite{Khandai:2013gva} \cite{Wong:2013sca} that the functional form 
of the Tsallis distribution with thermodynamic origin in terms of the parameter 
$q$ is the same as the QCD-inspired Hagedorn formula in terms of the parameter 
$n$~\cite{Hagedorn:1983wk} \cite{Blankenbecler:1974tm}. The parameter $n$ or 
$q$ governs the power law tail of the $p_T$ spectra. The Tsallis function gives 
an excellent description of $p_{T}$ spectra of all identified mesons measured in 
pp collisions at RHIC and LHC energies \cite{Khandai:2013gva}\cite{Adare:2010fe}
\cite{Sett:2014csa}. 
The hadron $p_T$ distributions in heavy ion collisions are modified due to 
collective flow so the Tsallis blast wave method is used as in 
Ref.~\cite{Tang:2008ud}. The average transverse flow is included in the Tsallis 
distribution in Ref.~\cite{Khandai:2013fwa} keeping the functional form analytical. 
This modified function can be used in a wider $p_T$ range as was done for both 
meson and baryon spectra in heavy ion collisions at RHIC and LHC 
~\cite{Khandai:2013fwa} \cite{Sett:2015lja}. A Tsallis analysis of $p_{T}$ spectra 
of identified particles produced in pp collisions at various RHIC and LHC energies 
has been performed in Ref. \cite{Zheng:2015tua}.

The ALICE experiment has measured identified charged hadron spectra  in PbPb 
collisions at $\sqrt{s_{NN}}$ = 2.76 TeV for many centrality classes~
\cite{Abelev:2013vea}. They have also carried out a combined blast wave analysis 
of all particle spectra in a single freeze-out scenario choosing an intermediate 
$p_T$ range where hydrodynamics is assumed to be applicable. They obtained inverse 
slope parameter of blast wave function in chosen fit-ranges of particle spectra. 
Few analyses perform fits in terms of different sets of parameters for strange and 
non-strange particles \cite{Sett:2015lja}.  There are recent analyses of LHC data 
\cite{Thakur:2016boy} \cite{Lao:2015zgd} \cite{Wei:2016ihj} suggesting a 
differential freeze-out scenario for different particles. The Tsallis study of 
$p_T$ spectra in pp and AA collisions in Ref~\cite{Lao:2015zgd} finds heavier 
particles freeze out at higher temperatures. They also carried out a blast wave 
analysis to obtain radial flow of particles. 
The work in Refs.~\cite{Lao:2015zgd}\cite{Wei:2016ihj} analyses the $p_{T}$ spectra 
of $\pi^{+}, K^{+}, p, d$ and $^{3}He$ particles in PbPb collisions at 
$\sqrt{s_{NN}}$ = 2.76 TeV using the Tsallis distribution in different centralities. 
The extracted temperature is found to increase with the increase of the particle 
rest mass. The intercept in temperature and mass correlation has been regarded as 
the mean kinetic freeze out temperature of the system. 
The higher effective temperature in the central collisions as compared to that 
in the peripheral collisions is attributed to the higher excitation state of the 
system in central collisions at the time of freeze-out \cite{Wei:2016ihj}. 
The transverse flow velocity of the produced particles in the source rest frame 
is extracted based on the slopes in the linear relation between the mean $p_{T}$ 
and mean moving mass.

In this work, we use the Tsallis distribution with transverse flow to carry 
out the analysis of $p_{T}$ spectra of strange hadrons in different multiplicity 
events produced in pp collision at $\sqrt{s}$ = 7 TeV, pPb collision at 
$\sqrt{s_{NN}}$ = 5.02 TeV and PbPb collision at $\sqrt{s_{NN}}$ = 2.76 TeV. 
We perform the fits of spectra of strange hadrons $K^0_s$, $\Lambda$ and $\Xi^-$ 
considering both single and differential freeze out scenarios. 
The $(p_{T})$ distributions of these hadrons in different systems are characterized 
in terms of the parameters namely, Tsallis temperature $(T)$, power $(n)$ and 
average radial flow velocity $(\beta)$. The goal is to study the behaviour of the 
three systems with two different freeze out scenarios.

\section{Tsallis Distribution Function with Transverse flow} 

The transverse momentum spectra of hadrons can be described using the modified 
Tsallis distribution \cite{Khandai:2013fwa}. The modified Tsallis function is 
given by 
\begin{equation}
E \frac{d^{3}N}{dp^{3}} = C_{n} \Bigg[\exp\Bigg(\frac{- \gamma ~ 
\beta ~ p_{T}}{n ~ T}\Bigg) + \frac{\gamma ~ m_{T}}{ n ~ T}\Bigg]^{-n}~.
\label{pbpb_equation1one}
\end{equation}
Here $C_{n}$ is the normalization constant, $m_{T} (\sqrt{p^{2}_{T} + m^{2}})$ is 
the transverse mass, $\gamma = 1/\sqrt{ 1 - \beta^{2}}$, $\beta$ is the average 
transverse velocity of the system and $T$ is the temperature. The power $n$ is 
related to the Tsallis parameter $q$ as $n = 1/(q - 1)$, where $q$ gives temperature 
fluctuations \cite{Wilk:1999dr} in the system as : 
$q - 1$ = Var($T$)/$ \textless~ T~ \textgreater^{2}$. The parameter $n$ can be 
called degree of thermalization \cite{Wilk:1999dr}. Larger values of $n$ correspond 
to smaller values of $q$. Both $n$ and $q$ have been interchangeably used in the 
Tsallis distribution \cite{Biro:2008hz} \cite{Adare:2010fe} \cite{Cleymans:2012ya} 
\cite{Adare:2011vy} \cite{Abelev:2006cs}. The Tsallis interpretation of parameters 
$T$ as temperature and $q$ as non-extensivity parameter is more suited for heavy 
ion collisions while for pp collisions Hagegorn interpretation in terms of power 
$n$ and inverse slope parameter $T$ is more meaningful. Phenomenological studies 
suggest that, for quark-quark point scattering, $n~\sim$~4 
\cite{Blankenbecler:1975ct} \cite{Brodsky:2005fza}, and when multiple scattering 
centers are involved $n$ grows larger.

When $\beta$ is zero,  Eq. \ref{pbpb_equation1one} is the usual Tsallis equation : 
\begin{equation}
E \frac{d^{3} N}{d p^{3}} = C_{n}~ \Bigg(1 + \frac{m_{T}}{n T}\Bigg)^{- n}~.
\label{pbpb_equation2two}
\end{equation}
In recent years, Tsallis distribution has been the most popular tool to characterize 
hadronic collisions \cite{Wong:2013sca} \cite{Wilk:1999dr} \cite{Cleymans:2012ya} 
\cite{Cleymans:2008mt}. At low $p_{T}$, Eq. \ref{pbpb_equation1one} represents a 
thermalized system with collective flow 

\begin{equation}
E \frac{d^{3} N}{d p^{3}} \simeq C_{n}~ \exp\Bigg(\frac{- \gamma (m_{T} 
- \beta p_{T})}{T}\Bigg)~ {\rm{for}} ~ p_{T} \rightarrow ~ 0~.
\label{pbpb_equation3three}
\end{equation}
At high $p_{T}$, it becomes a power law
\begin{equation}
  E \frac{d^{3} N}{d p^{3}} \simeq C_{n}~ \Bigg(\frac{\gamma m_{T}}{n T}\Bigg)^{- n}~ 
{\rm{for}} ~ p_{T} \rightarrow ~ \infty~~. 
\label{pbpb_equation4four}
\end{equation}
 For very large $n(\rightarrow \infty)$ (or $q \rightarrow$ 1), the 
Eq. \ref{pbpb_equation1one} takes the Boltzmann form with transverse flow as 
in Eq.~\ref{pbpb_equation3three}.

\section{Results and Discussions}

In the present analysis, we use $p_{T}$ spectra of the strange hadrons produced 
in different multiplicity events of pp collision at $\sqrt{s}$ = 7 TeV, pPb 
collision at $\sqrt{s_{NN}}$ = 5.02 TeV and PbPb collision at $\sqrt{s_{NN}}$ 
= 2.76 TeV measured by CMS experiment \cite{Khachatryan:2016yru}. 
The measured $p_{T}$ spectra of $K^{0}_{s}, \Lambda$ and $\Xi^{-}$ particles are 
fitted with the modified Tsallis distribution (Eq. \ref{pbpb_equation1one}) 
using two methods. First, we assume a differential freeze out scenario and analyse 
the $p_{T}$ spectra of $K^{0}_{s}, \Lambda$ and $\Xi^{-}$ individually. In the second 
method, we assume a single freeze out scenario and perform a combined fitting of 
all the three hadron spectra simultaneously in each multiplicity bin.

\begin{figure}
\resizebox{0.5\textwidth}{!}{
\includegraphics{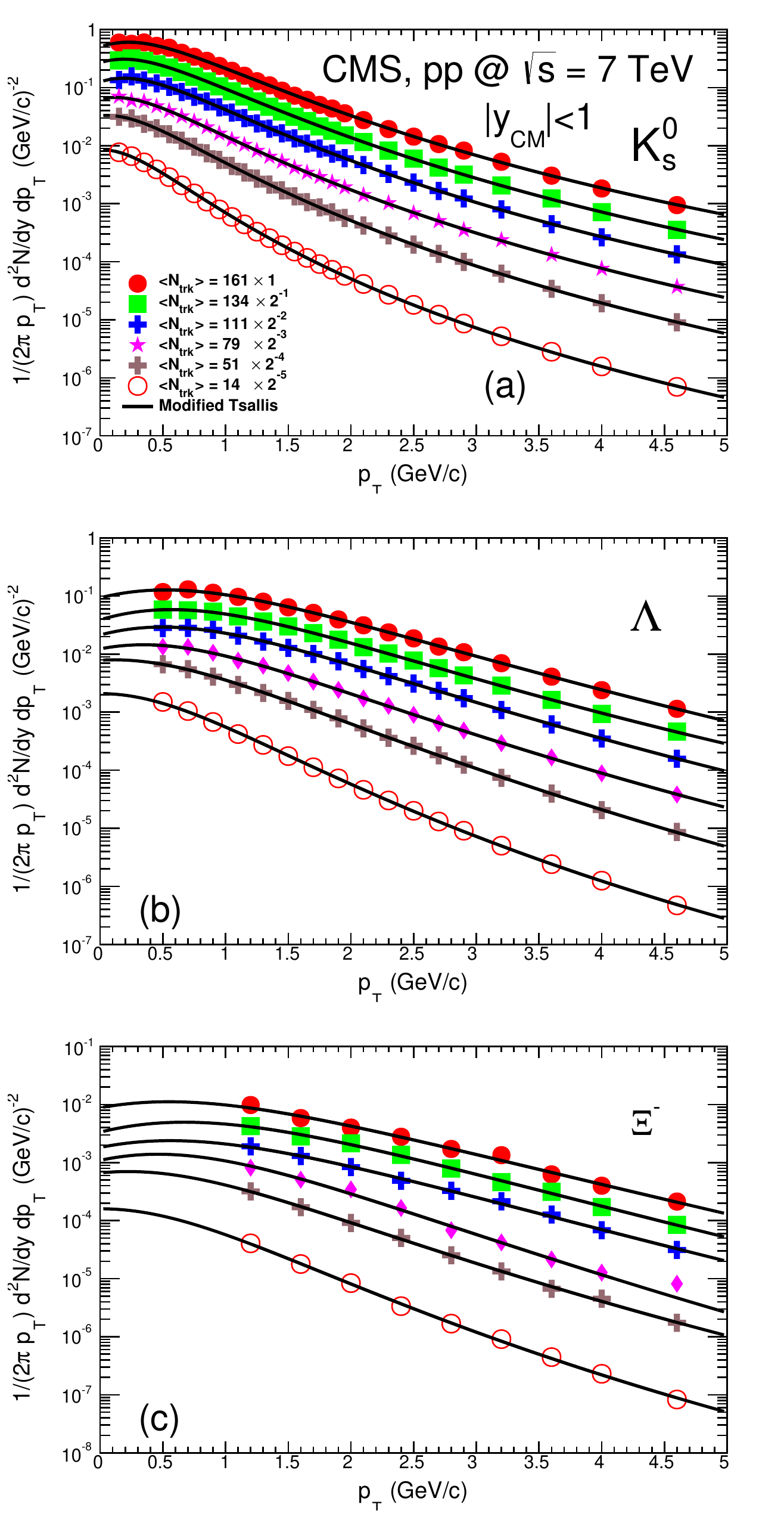}}
\caption{The invariant yields of the $K^{0}_{s}, \Lambda$ and $\Xi^{-}$ hadrons as a 
function of the transverse momentum $p_{T}$ for pp collisions at $\sqrt{s}$ = 7 TeV
measured by CMS \cite{Khachatryan:2016yru}. The yields are shown for different 
multiplicity bins. The solid curves are the fitted modified Tsallis distribution.}
\label{Figure1_pp_collision_7tev}
\end{figure}

Figure \ref{Figure1_pp_collision_7tev} shows the invariant yields of the strange \\
hadrons $\big((a)~ K^{0}_{s},~ (b)~ \Lambda~\rm{and}~ (c)~ \Xi^{-}\big)$ as a function 
of $p_{T}$ for pp collisions at $\sqrt{s}$ = 7 TeV measured by the CMS experiment 
\cite{Khachatryan:2016yru} in the mid rapidity, $|y_{\rm{CM}}|<1$. 
The invariant yields are given for six multiplicity classes which correspond to the 
efficiency corrected average track multiplicities 
$<N_{\rm trk}>$ = 14, 51, 79, 111, 134 and 161. 
The solid curves are the modified Tsallis distributions fitted individually to 
different hadron spectra. 
The individual fitting gives excellent fit quality for all the multiplicity classes 
which can be inferred from the values of $\chi^{2}/\rm{NDF}$ given in the table 
\ref{individual_fitting_at_7tev_pp_collision_chi2_NDF}.

\begin{figure}
\resizebox{0.40\textwidth}{!}{
\includegraphics{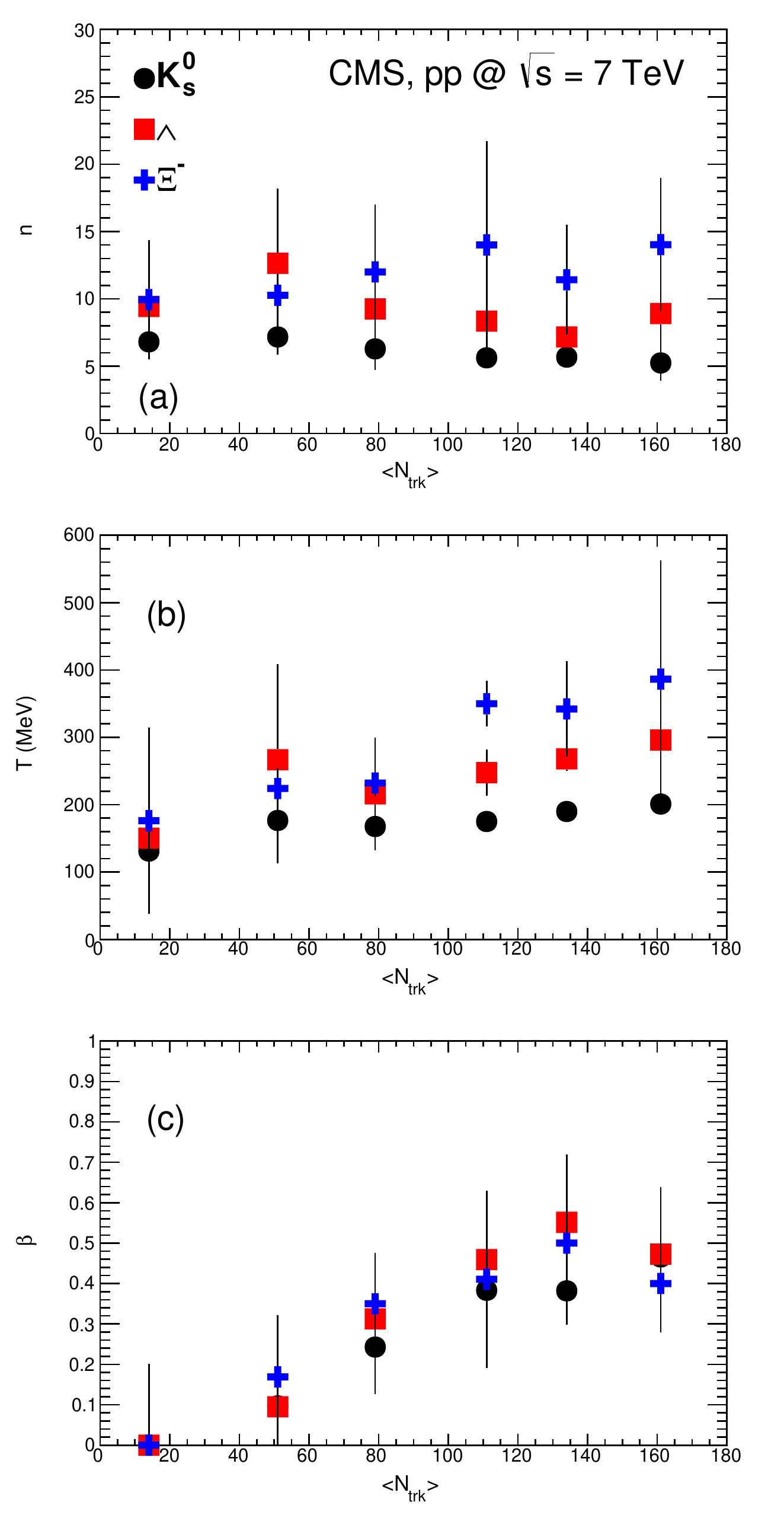}}
\caption{The Tsallis parameters $n, T$ and $\beta$ for the strange hadrons as 
a function of mean track multiplicity $<N_{\rm{\rm trk}}>$ of event class in pp 
collision at $\sqrt{s}$ = 7 TeV.}
\label{Figure2_pp_collision_7tev}
\end{figure}

Figure \ref{Figure2_pp_collision_7tev} (a) shows the Tsallis parameter $n$ for 
the strange hadrons $K^{0}_{s},~ \Lambda~\rm{and}~ \Xi^{-}$ as a function of event 
multiplicity in the pp collisions at $\sqrt{s}$ = 7 TeV. The parameters $T$ and 
$\beta$ are shown in panels (b) and (c) respectively.
The value of $n$ increases with the mass of particles but there is little variation 
with multiplicity. It means that the degree of thermalization remains similar for 
the events of different multiplicity classes. The value of Tsallis temperature $T$ 
increases with the multiplicity for all three hadrons and becomes large for the 
highest multiplicity events.  
The transverse flow also increases with multiplicity for all the three hadrons. 
The highest value of $\beta$ remains in the range 0.4-0.5 in pp system.

\begin{figure}
\resizebox{0.5\textwidth}{!}{
\includegraphics{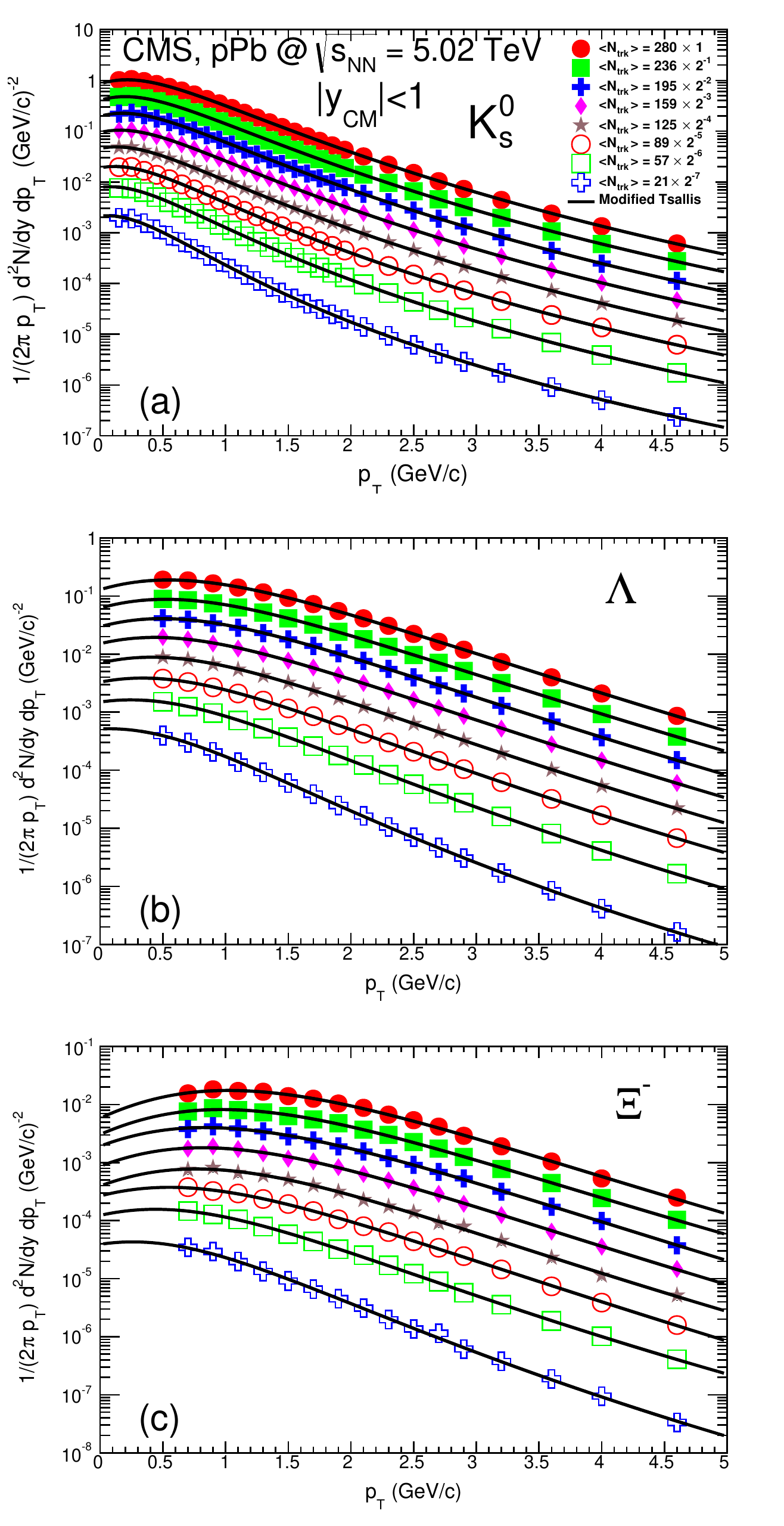}}
\caption{The invariant yields of the $K^{0}_{s}$, $\Lambda$ and $\Xi^{-}$ hadrons 
as a function of the transverse momentum $p_{T}$ for pPb collisions at 
$\sqrt{s_{NN}}$ = 5.02 TeV measured by CMS \cite{Khachatryan:2016yru}. The yields 
are shown for different multiplicity bins. The solid curves are the fitted 
modified Tsallis distribution.}
\label{Figure3_pPb_502tev}
\end{figure}

Figure \ref{Figure3_pPb_502tev} shows the invariant yields of $K^{0}_{s}$, $\Lambda$ 
and $\Xi^{-}$ as a function of $p_{T}$ in panels (a), (b) and (c) respectively
for pPb collisions at $\sqrt{s_{NN}}$ = 5.02 TeV measured by the CMS experiment 
\cite{Khachatryan:2016yru}. 
The invariant yields are given for eight multiplicity classes which correspond 
to the efficiency corrected average track multiplicities 
$<N_{\rm trk}>$ = 21, 57, 89, 125, 159, 195, 236 and 280 \cite{Chatrchyan:2013nka}. 
The solid curves are the modified Tsallis distributions fitted individually to 
different hadron spectra. The individual fitting gives excellent fit quality for 
all the multiplicity classes which can be seen from the values of $\chi^{2}/\rm{NDF}$ 
given in the table \ref{individual_fitting_at_502tev_pPb_collision_chi2_NDF}. 

\begin{figure}
\resizebox{0.4\textwidth}{!}{
\includegraphics{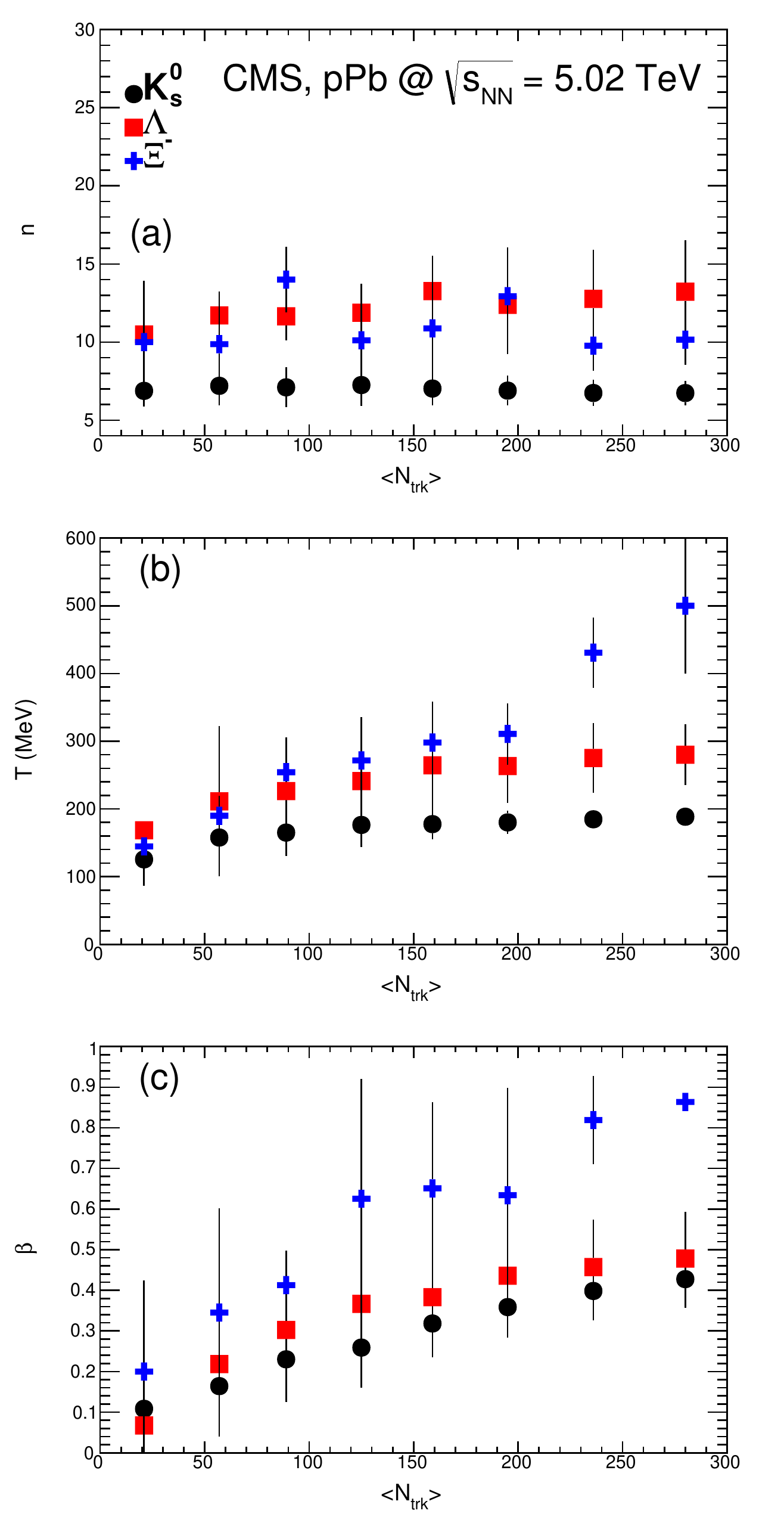}}
\caption{The Tsallis parameters $n, T$ and $\beta$ for the strange hadrons as a 
function of mean track multiplicity $<N_{\rm{\rm trk}}>$ in pPb collision at 
$\sqrt{s_{NN}}$ = 5.02 TeV.}
\label{Figure4_pPb_502tev}
\end{figure}

Figure \ref{Figure4_pPb_502tev} (a) shows the Tsallis parameter $n$ for the strange 
hadrons $K^{0}_{s}$, $\Lambda$ and $\Xi^{-}$ as a function of event multiplicity in 
the pPb collisions at $\sqrt{s_{NN}}$ = 5.02 TeV. The parameters $T$ and $\beta$ 
are shown in panels (b) and (c) respectively. 
Like pp system, similar conclusions can be drawn for the pPb system. The degree of 
thermalization remains similar for the events of different multiplicity classes. 
The value of Tsallis temperature $T$ increases with the multiplicity for all three 
hadrons and becomes large for the highest multiplicity events.  
The transverse flow also increases with multiplicity for all the three hadrons. 
The highest value of $\beta$ remains in the range 0.4-0.5 for 
$K^{0}_{s}~ \rm{and}~ \Lambda$ and 0.75-0.85 for $\Xi^{-}$ in pPb system.

\begin{figure}
\resizebox{0.5\textwidth}{!}{
\includegraphics{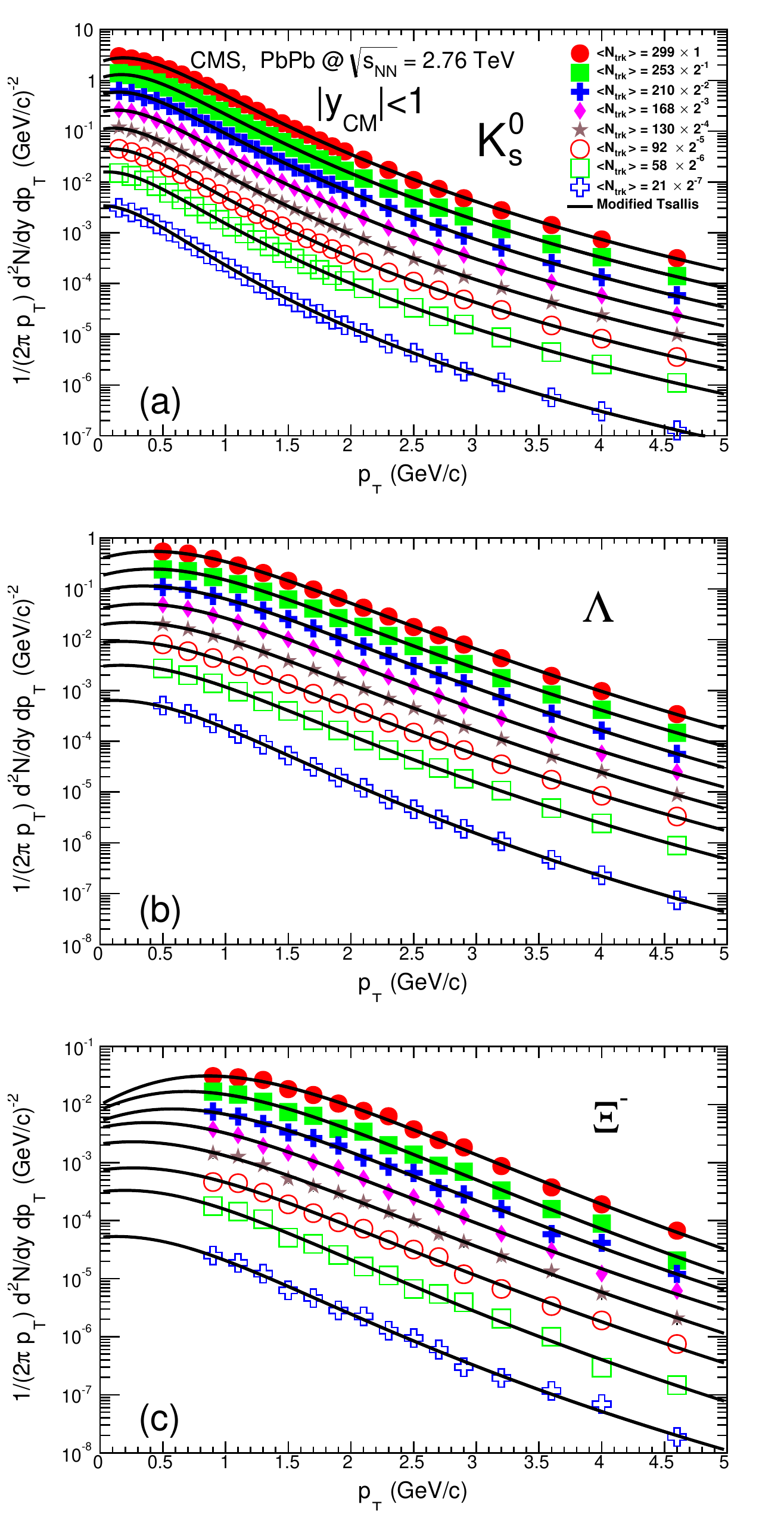}}
\caption{The invariant yields of the $K^{0}_{s}$, $\Lambda$ and $\Xi^{-}$ hadrons as 
a function of the transverse momentum $p_{T}$ for PbPb collisions at $\sqrt{s_{NN}}$ 
= 2.76 TeV measured by CMS \cite{Khachatryan:2016yru}. The yields are shown for 
different multiplicity bins. The solid curves are the fitted modified Tsallis 
distribution.}
\label{Figure5_PbPb_276tev}
\end{figure}

Figure \ref{Figure5_PbPb_276tev} shows the invariant yields of $K^{0}_{s}$, $\Lambda$ 
and $\Xi^{-}$ as a function of $p_{T}$ in panels (a), (b) and (c) respectively for 
PbPb collisions at $\sqrt{s_{NN}}$ = 2.76 TeV measured by the CMS experiment 
\cite{Khachatryan:2016yru}. 
The invariant yields are given for eight multiplicity classes which correspond 
to the average efficiency corrected track multiplicities $<N_{\rm trk}>$ = 21, 58, 
92, 130, 168, 210, 253 and 299 \cite{Chatrchyan:2013nka}. The solid curves are 
the modified Tsallis distributions fitted individually to different hadron spectra. 
The individual fitting gives excellent fit quality for all the multiplicity classes 
which can be seen from the values of $\chi^{2}/\rm{NDF}$ given in the table 
\ref{individual_fitting_at_276tev_PbPb_collision_chi2_NDF}.

\begin{figure}
\resizebox{0.4\textwidth}{!}{
\includegraphics{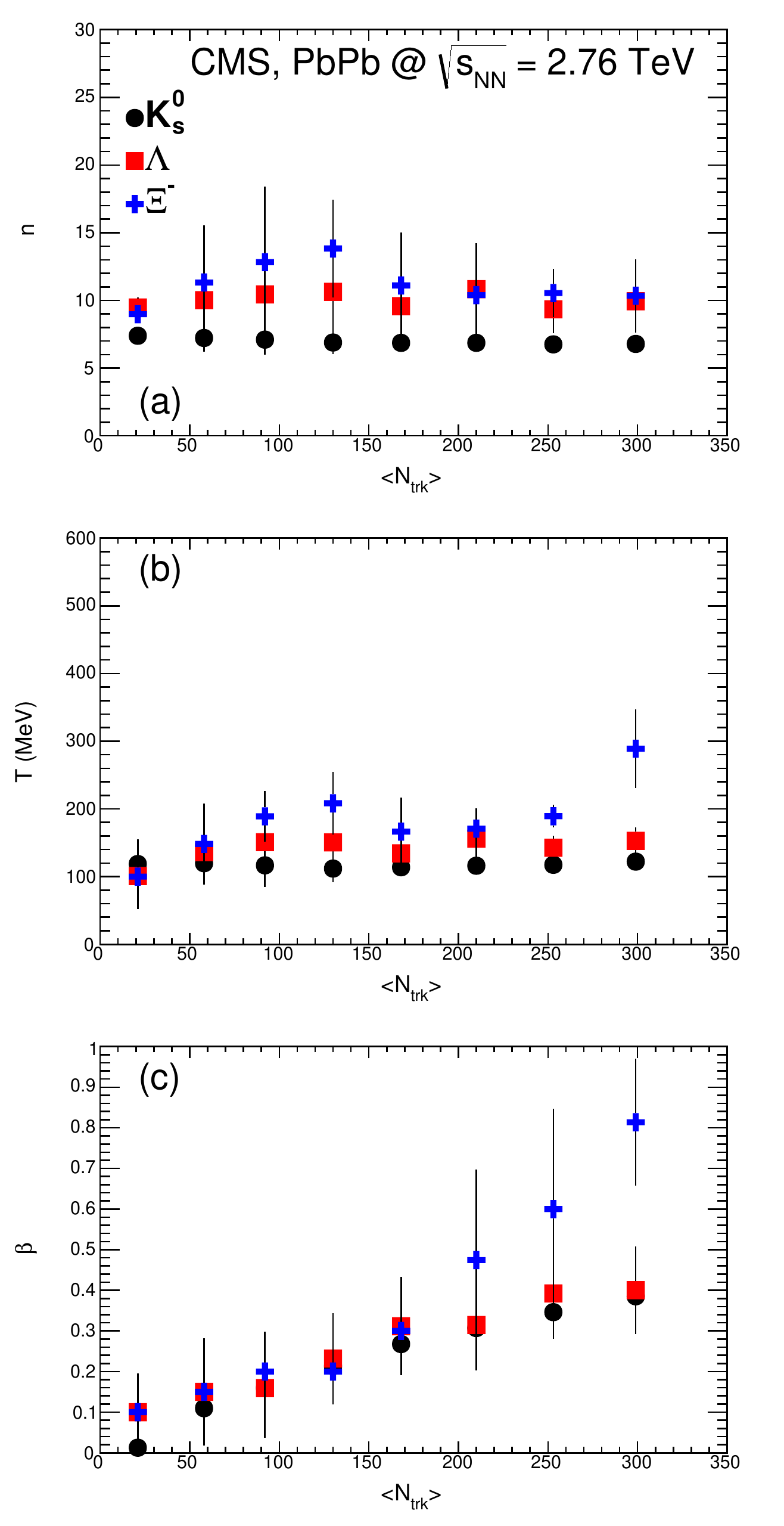}}
\caption{The Tsallis parameters $n, T$ and $\beta$ for the strange hadrons as a 
function of mean track multiplicity $<N_{\rm{\rm trk}}>$ in PbPb collision at 
$\sqrt{s_{NN}}$ = 2.76 TeV.}
\label{Figure6_PbPb_276tev}
\end{figure}

Figure \ref{Figure6_PbPb_276tev} (a) shows the Tsallis parameter $n$ for the strange 
hadrons $K^{0}_{s},~ \Lambda~\rm{and}~ \Xi^{-}$ as a function of event multiplicity 
in the PbPb collisions at $\sqrt{s_{NN}}$ = 2.76 TeV.  The parameters $T$ and $\beta$ 
are shown in panels (b) and (c) respectively. The degree of thermalization remains 
similar for the events of different multiplicity classes of PbPb collisions. Unlike 
pp and pPb systems, the value of Tsallis temperature $T$ varies little with the 
multiplicity for all three hadrons. 
The transverse flow $\beta$ also increases with multiplicity for all the three 
hadrons. The highest value of $\beta$ remains in the range 0.3-0.4 for 
$K^{0}_{s}~ \rm{and}~ \Lambda$ and 0.75-0.85 for $\Xi^{-}$ in PbPb system.

We have studied the spectra in the three systems pp, pPb and PbPb. Since all the 
three systems are at different collision energy, they cannot be compared directly. 
Moreover, how the multiplicity class can be related to the system size is not clear. 
For PbPb system, the highest multiplicity class presented corresponds to a centrality 
class of 55 $\%$, which can be called semi-peripheral events. 
It is still possible to draw some general comments from the present analysis. The 
behaviour of various parameters like $n$ and $\beta$ with the event multiplicity 
are similar for the three systems. 
The Tsallis temperature $T$ increases with the multiplicity for pp and pPb systems 
but it does not show a noticeable change with the event multiplicity for the PbPb 
system. 
The PbPb system has smaller temperature and transverse flow as compared to the pp 
and pPb systems. The PbPb system being large will maintain collectivity for longer
time and hence freeze-out at lower temperature.

We also carry out the above study using a single freeze out scenario for all the 
hadrons. It is possible to find out same set of parameters correspond to a good 
fit using combined fitting of the three hadron spectra.  
The combined fitting is performed in the $p_T$ range 0.3-3.8 GeV for $K^{0}_{s}$, 
0.6-3.8 GeV for $\Lambda$ and 1.2-3.8 GeV for $\Xi^{-}$. 
The $p_T$ range for all the particle spectra is kept same over the three systems.
   
\begin{figure}
\resizebox{0.5\textwidth}{!}{
\includegraphics{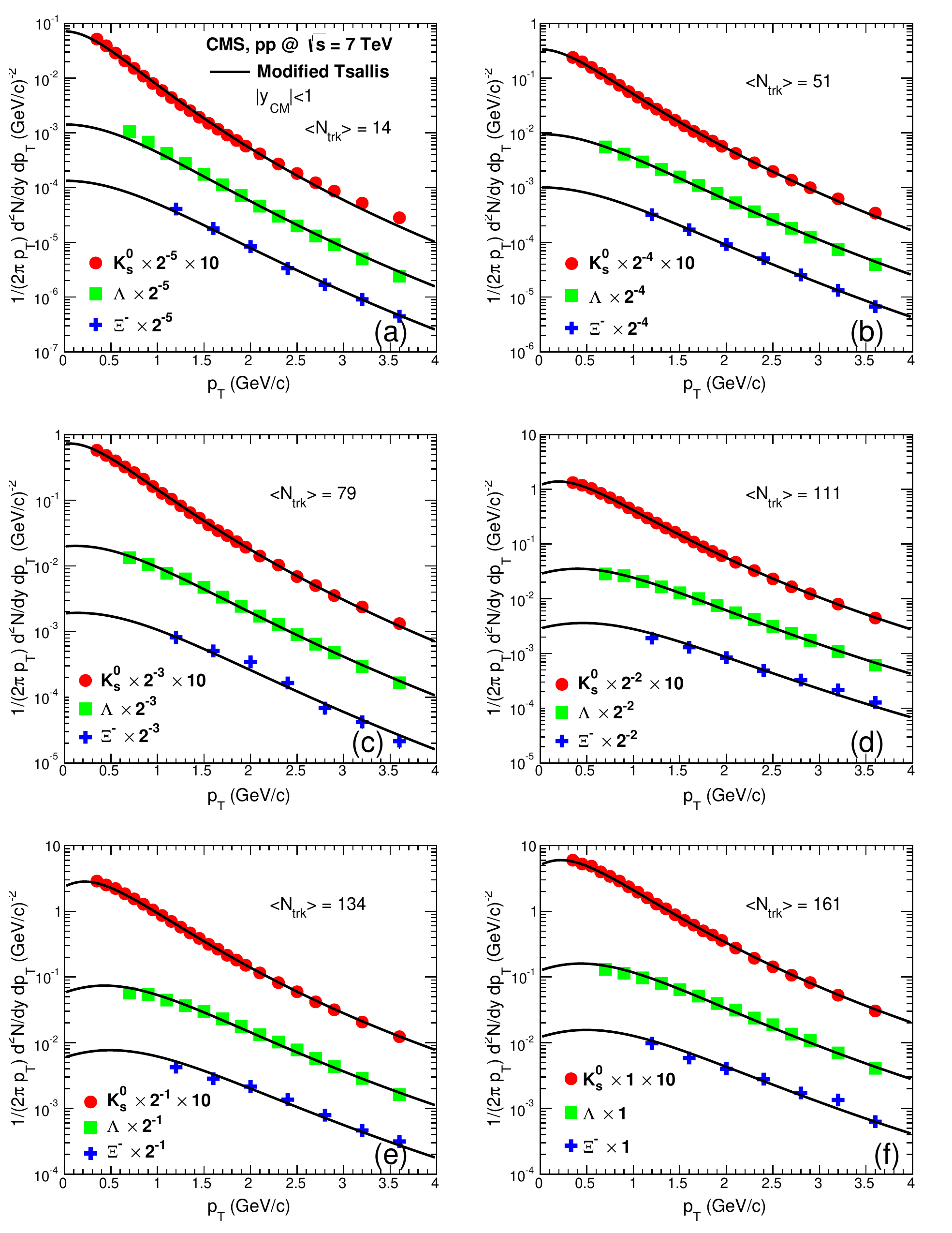}}
\caption{The invariant yields of the $K^{0}_{s}$, $\Lambda$ and $\Xi^{-}$ hadrons 
as a function of the transverse momentum $(p_{T})$ for pp collisions at 
$\sqrt{s}$ = 7 TeV measured by CMS \cite{Khachatryan:2016yru}. Each panel 
corresponds to a different event multiplicity class. The solid curves are the 
fitted modified Tsallis distribution.}
\label{Figure7_pp_collision_7tev}
\end{figure}

Figure \ref{Figure7_pp_collision_7tev} shows the invariant yields of strange 
hadrons $K^{0}_{s}$ (red circle), $\Lambda$ (green square) and $\Xi^{-}$ (blue cross) 
as a function of $p_{T}$ for pp collisions at $\sqrt{s}$ = 7 TeV measured by CMS 
experiment \cite{Khachatryan:2016yru} for average track multiplicities 
$<N_{\rm trk}>$ = 14 (panel a) , 51 (panel b), 79 (panel c), 111 (panel d), 
134 (panel e) and 161 (panel f). 
The solid curves are the modified Tsallis distributions fitted simultaneously to 
the three hadron spectra for a given multiplicity class. While the fit quality in 
individual fitting was better, the combined fitting gives reasonable quality fits 
for all multiplicity classes (except for the lowest multiplicity class) which can 
be judged from the values of $\chi^{2}/\rm{NDF}$ given in the table 
\ref{strange_particle_production_PbPb_collision_chi2_NDF}. 

\begin{figure}
\resizebox{0.40\textwidth}{!}{
\includegraphics{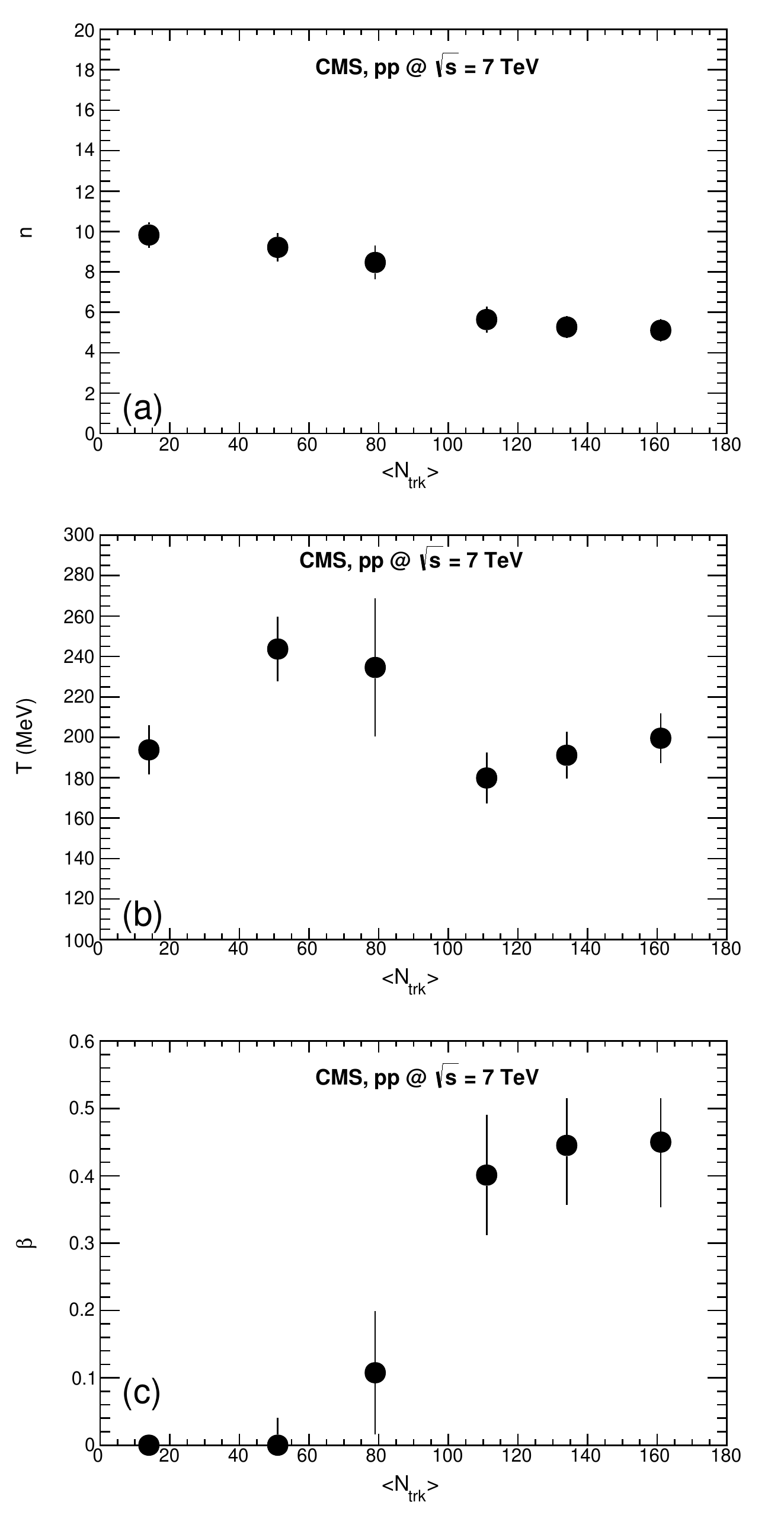}}
\caption{The Tsallis parameters $n, T$ and $\beta$ for the strange hadrons 
as a function of mean track multiplicity $<N_{\rm{\rm trk}}>$ in pp collision 
at $\sqrt{s}$ = 7 TeV.}
\label{Figure8_pp_collision_7tev}
\end{figure}

Figure \ref{Figure8_pp_collision_7tev} (a) shows the Tsallis parameter $n$ obtained 
from combined fitting of three strange hadrons as a function of the event 
multiplicity in the pp collision at $\sqrt{s}$ = 7 TeV. The parameters $T$ and 
$\beta$ are shown in panels (b) and (c) respectively. 
It can be seen from the figure that $n$ decreases with multiplicity and its value 
becomes $\sim 5$ for the high multiplicity classes.
The temperature $T$ does not show any trend with increasing multiplicity but it 
increases smoothly in the three highest multiplicity bins. The transverse flow 
$\beta$ is zero for the first two multiplicity bins and then sharply increases 
up to 0.5 in the three highest multiplicity bins.

\begin{figure}
\resizebox{0.5\textwidth}{!}{
\includegraphics{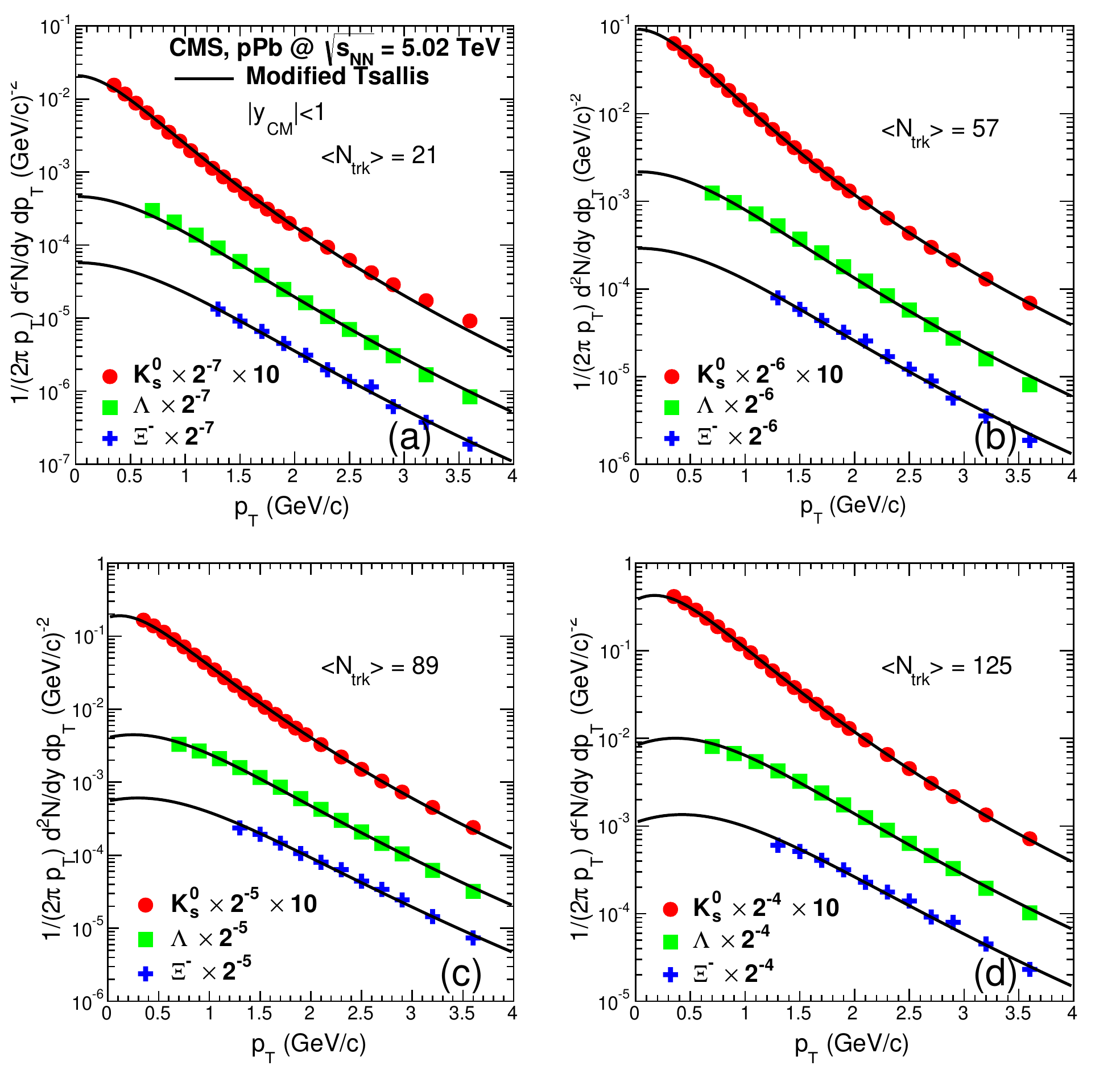}}
\caption{The invariant yields of the $K^{0}_{s}$, $\Lambda$ and $\Xi^{-}$ hadrons 
as a function of the transverse momentum $p_{T}$ for pPb collisions at 
$\sqrt{s_{NN}}$ = 5.02 TeV measured by CMS \cite{Khachatryan:2016yru}. Different 
panels correspond to different multiplicity bins. 
The solid curves are the fitted modified Tsallis distribution.}
\label{Figure9_pPb_502tev}
\end{figure}

\begin{figure}
\resizebox{0.5\textwidth}{!}{
\includegraphics{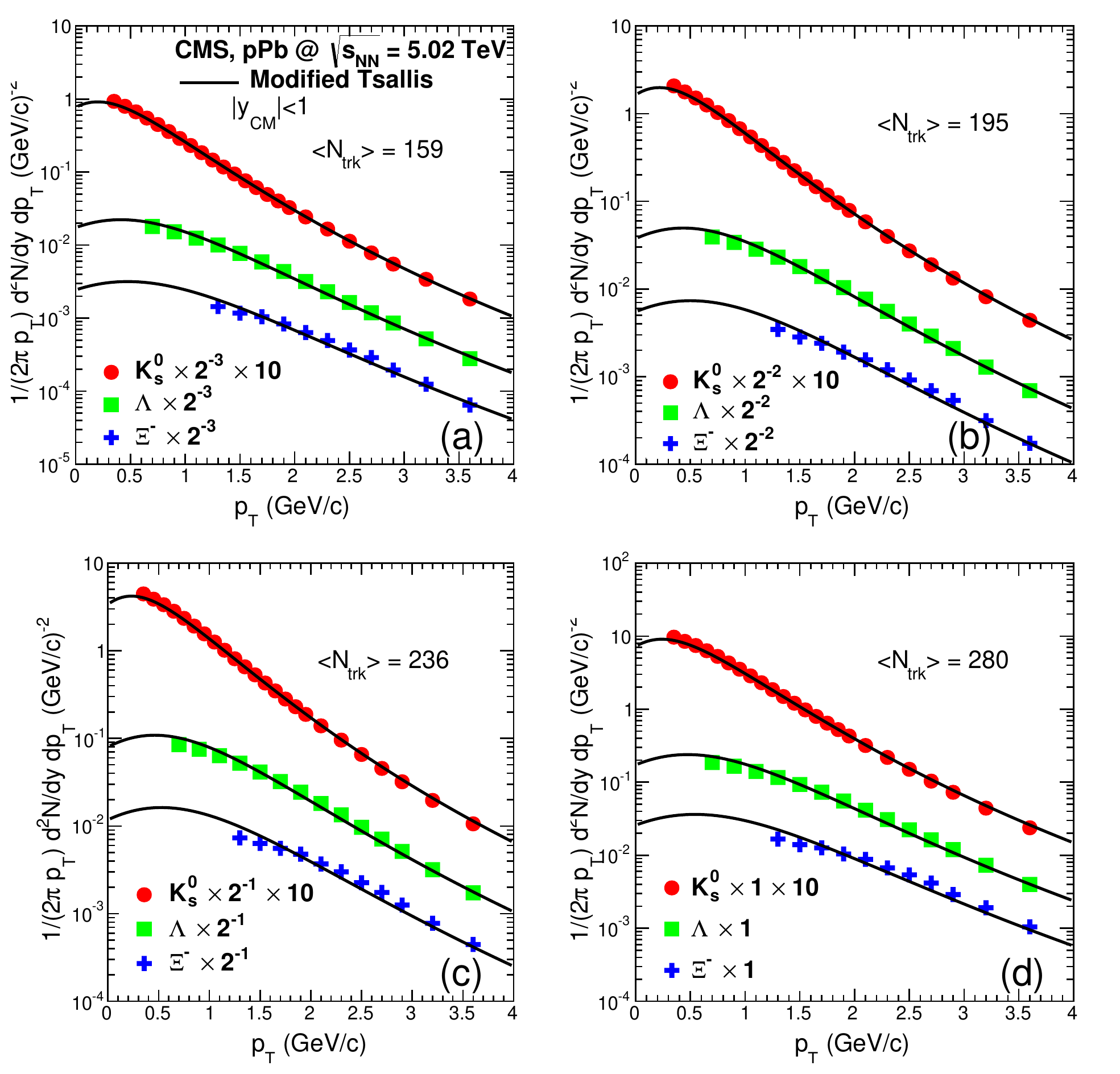}}
\caption{The invariant yields of the $K^{0}_{s}$, $\Lambda$ and $\Xi^{-}$ hadrons 
as a function of the transverse momentum $p_{T}$ for pPb collisions at 
$\sqrt{s_{NN}}$= 5.02 TeV measured by CMS \cite{Khachatryan:2016yru}. Different 
panels correspond to different multiplicity bins. 
 The solid curves are the fitted modified Tsallis distribution.}
\label{Figure10_pPb_502tev}
\end{figure}

Figure \ref{Figure9_pPb_502tev} shows the invariant yields of strange hadrons 
$K^{0}_{s}$, $\Lambda$ and $\Xi^{-}$ as a function of $p_{T}$ for pPb collisions 
at $\sqrt{s_{NN}}$ = 5.02 TeV measured by CMS experiment \cite{Khachatryan:2016yru} 
for the average track multiplicities 
$<N_{\rm trk}>$ = 21 (panel a) , 57 (panel b), 89 (panel c) and 125 (panel d). 
Figure \ref{Figure10_pPb_502tev} corresponds to average track multiplicities
$<N_{\rm trk}>$ = 159 (panel a) , 195 (panel b), 236 (panel c) and 280 (panel d) 
in pPb collisions. 
In both the figures,  the solid curves are the modified Tsallis distributions 
fitted simultaneously to the three strange hadrons. 
The fit quality is reasonably good except for the two lowest multiplicity classes 
which can be judged from the values of $\chi^{2}/\rm{NDF}$ given in the table 
\ref{strange_particle_production_PbPb_collision_chi2_NDF}.

\begin{figure}
\resizebox{0.4\textwidth}{!}{
\includegraphics{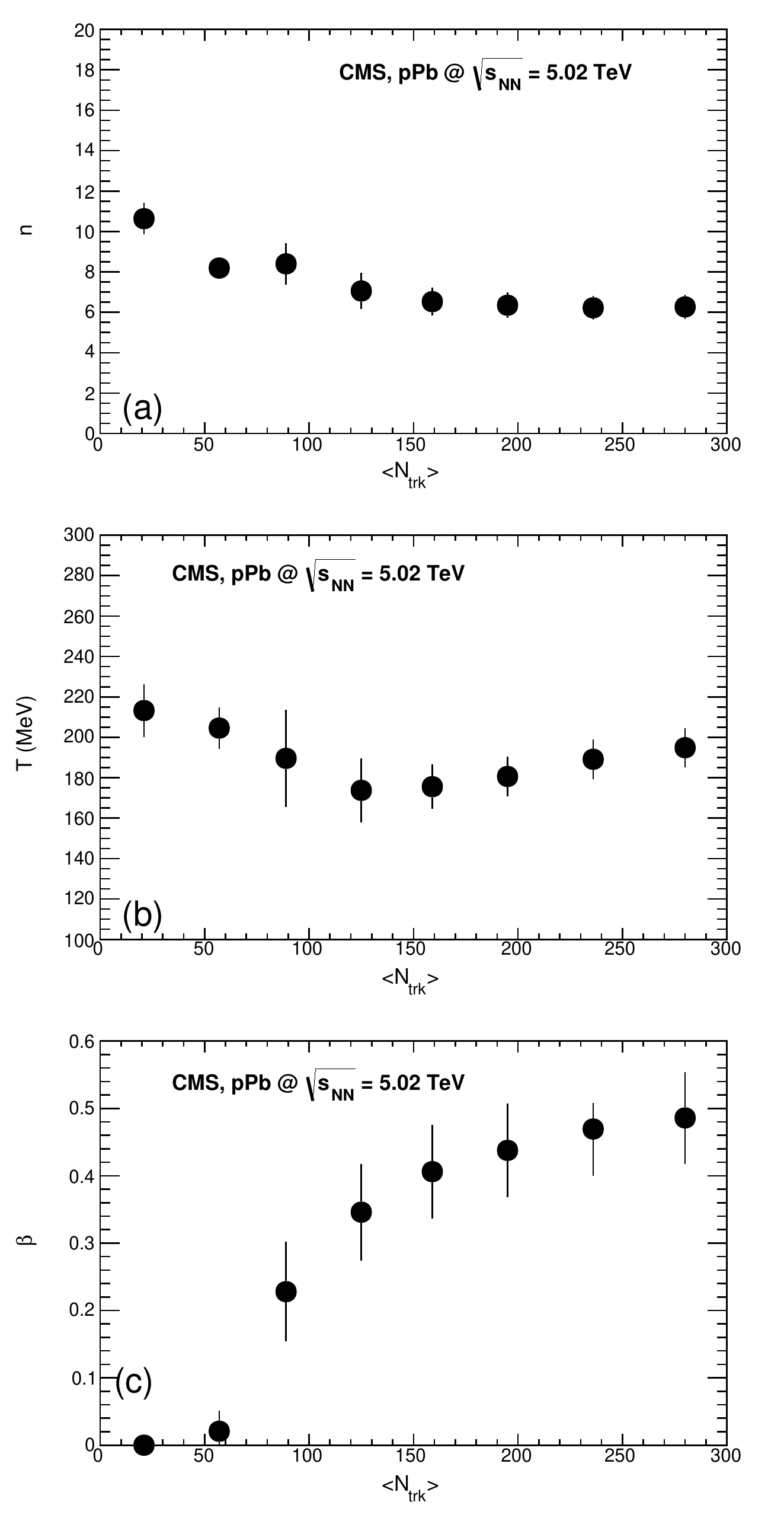}}
\caption{The Tsallis parameters $n, T$ and $\beta$ for the strange hadrons as 
a function of mean track multiplicity $<N_{\rm{\rm trk}}>$ in pPb collision at 
$\sqrt{s_{NN}}$ = 5.02 TeV.}
\label{Figure11_pPb_502tev}
\end{figure}

Figure \ref{Figure11_pPb_502tev} (a) shows the Tsallis parameter $n$ obtained from
combined fitting of three strange hadrons as a function of the event multiplicity 
in the pPb collision at $\sqrt{s_{NN}}$ = 5.02 TeV. The parameters $T$ and $\beta$ 
are shown in panels (b) and (c) respectively.
Like pp system, in pPb system also $n$ decreases with multiplicity and its value 
becomes $\sim 6$ for the high multiplicity classes. The temperature $T$ smoothly 
decreases with multiplicity for the first few bins then increases slightly in the 
last four most central bins. 
The transverse flow $\beta$ increases with the event multiplicity and roughly 
saturates at 0.5 for the largest multiplicity bins. 
On comparing figs. \ref{Figure8_pp_collision_7tev} and \ref{Figure11_pPb_502tev}
one can conclude that the values of parameters and their behaviour are the same 
if we pick up the same track classes of the pp and pPb systems.

\begin{figure}
\resizebox{0.5\textwidth}{!}{
\includegraphics{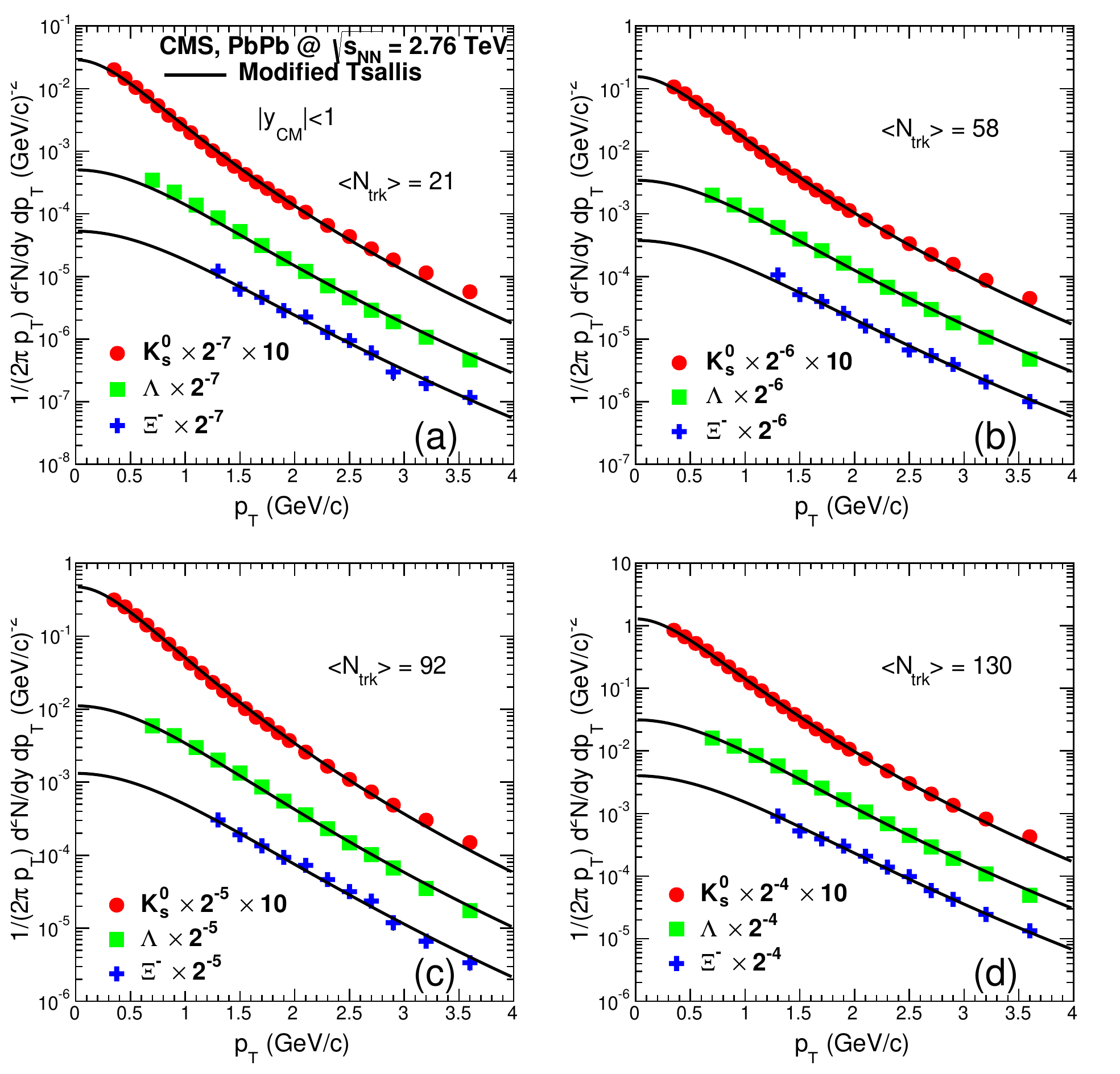}}
\caption{The invariant yields of the $K^{0}_{s}$, $\Lambda$ and $\Xi^{-}$ hadrons as 
a function of the transverse momentum $p_{T}$ for PbPb collisions at $\sqrt{s_{NN}}$ 
= 2.76 TeV measured by CMS \cite{Khachatryan:2016yru}. Different panels correspond 
to different multiplicity bins. 
The solid curves are the fitted modified Tsallis distribution.}
\label{Figure12_PbPb_276tev}
\end{figure}

\begin{figure}
\resizebox{0.5\textwidth}{!}{
\includegraphics{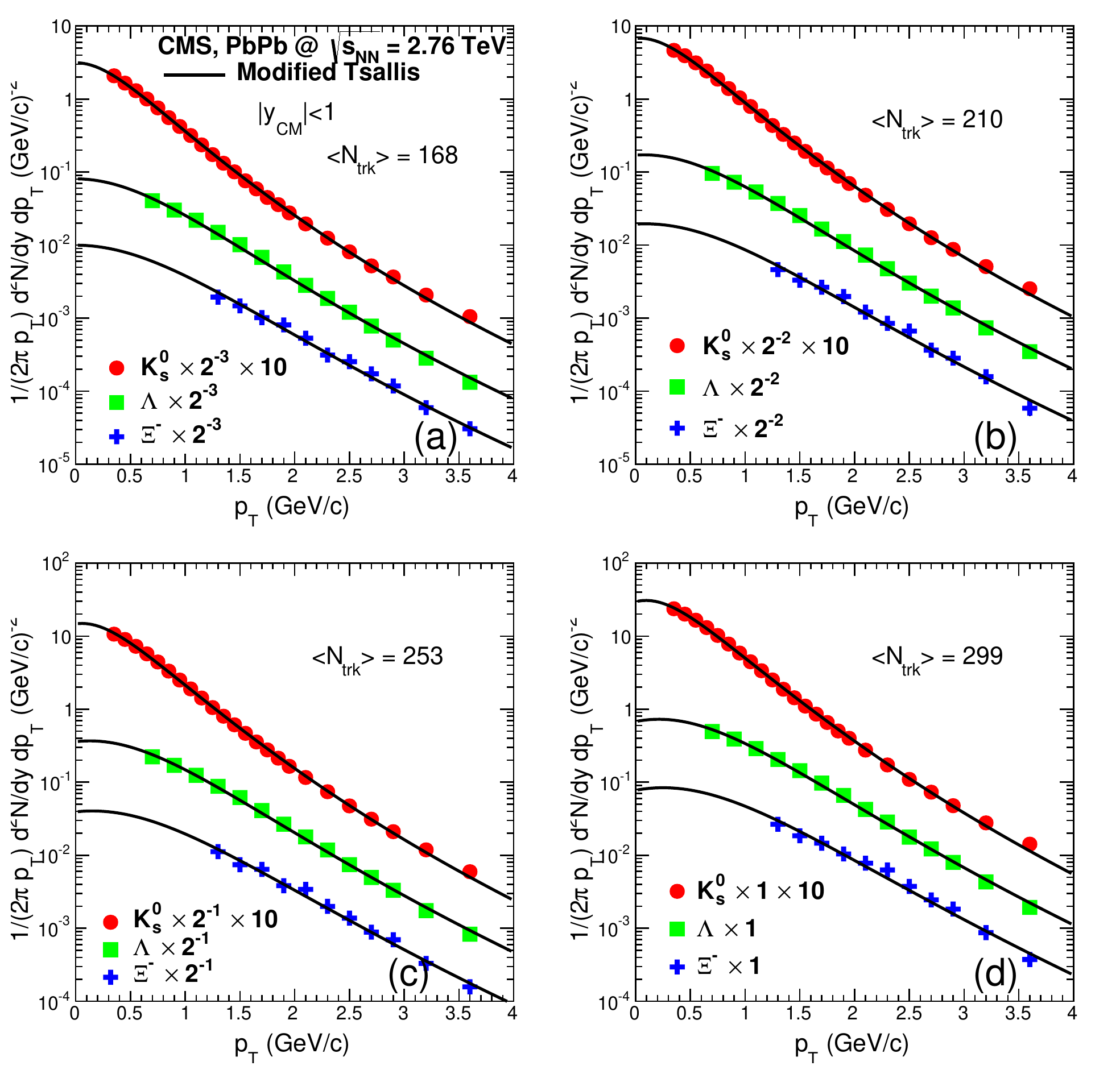}}
\caption{The invariant yields of the $K^{0}_{s}$, $\Lambda$ and $\Xi^{-}$ hadrons as 
a function of the transverse momentum $p_{T}$ for PbPb collisions at 
$\sqrt{s_{NN}}$ = 2.76 TeV measured by CMS \cite{Khachatryan:2016yru}. Different 
panels correspond to different multiplicity bins.
The solid curves are the fitted modified Tsallis distribution.}
\label{Figure13_PbPb_276tev}
\end{figure}

Figure \ref{Figure12_PbPb_276tev} shows the invariant yields of strange hadrons 
$K^{0}_{s}$, $\Lambda$ and $\Xi^{-}$ as a function of $p_{T}$ for PbPb collisions at 
$\sqrt{s_{NN}}$ = 2.76 TeV measured by CMS experiment \cite{Khachatryan:2016yru} for 
average track multiplicities $<N_{\rm trk}>$ = 21 (panel a), 58 (panel b), 92 (panel c) 
and 130 (panel d). 
Figure \ref{Figure13_PbPb_276tev} corresponds to average track multiplicities 
$<N_{\rm trk}>$ = 168 (panel a), 210 (panel b), 253 (panel c) and 299 (panel d).
In both the figures the solid curves are the modified Tsallis distributions fitted 
simultaneously to three strange hadrons.
The fit quality is reasonably good except for the lowest multiplicity classes 
which can be judged from the values of $\chi^{2}/\rm{NDF}$ given in the table 
\ref{strange_particle_production_PbPb_collision_chi2_NDF}.

\begin{figure}
\resizebox{0.4\textwidth}{!}{
\includegraphics{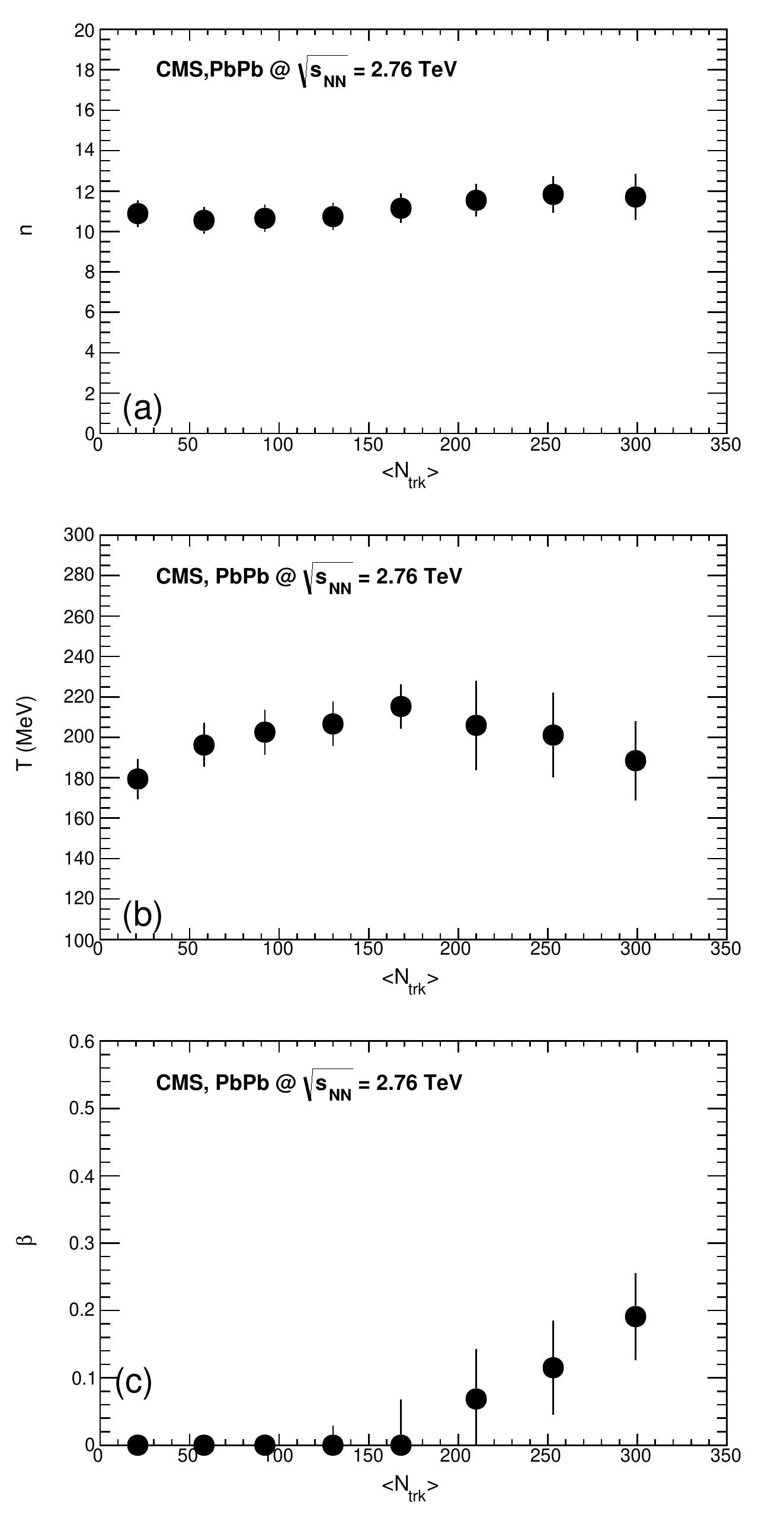}}
\caption{The Tsallis parameters $n, T$ and $\beta$ for the strange hadrons as a 
function of mean track multiplicity $<N_{\rm{\rm trk}}>$ in PbPb collision at 
$\sqrt{s_{NN}}$ = 2.76 TeV.}
\label{Figure14_PbPb_276tev}
\end{figure}

Figure \ref{Figure14_PbPb_276tev} (a) shows the Tsallis parameter $n$ obtained from
combined fitting of three strange hadrons as a function of the event multiplicity in 
the PbPb collision at $\sqrt{s_{NN}}$ = 2.76 TeV. The parameters $T$ and $\beta$ are 
shown in panels (b) and (c) respectively. 
The PbPb system looks different from the pp and pPb systems. The value of parameter 
$n$ shows a slight increase with the multiplicity and has a higher value ($\sim 12$) 
as compared to that for smaller systems. 
The temperature $T$ smoothly increases with multiplicity for the first few bins then 
decreases smoothly in the last four most central bins which is a trend opposite to 
what is found in the pPb collisions. The transverse flow velocity $\beta$ remains 
zero for the first five multiplicity classes and increases up to 0.2 for the highest 
multiplicity class. The transverse flow in PbPb system remains lower as compared 
to pp and pPb systems.
The highest multiplicity class of the PbPb system corresponds to a semi-peripheral 
collision and we expect that for central collisions the value of $\beta$ will become 
large.

\begin{table}[ht]
\caption{$\chi^{2}/\rm{NDF}$ of the strange hadrons production in pp collision at 
$\sqrt{s}$ = 7 TeV obtained using individual fitting}
\begin{center}
\begin{tabular}{| c || c | c | c |} 
\hline
&\multicolumn{3}{l|}{~~~~~~$\chi^{2}/\rm{NDF}$} \\ \hline 
$<N_{\rm{\rm trk}}>$ &  $K^{0}_{s}$    &  $\Lambda$     &  $\Xi^{-}$  \\    \hline\hline 
    14        &  0.022        &  0.106         &  0.198          \\ \hline 
    51        &  0.013        &  0.062         &  0.150        \\ \hline
    79        &  0.049        &  0.092         &  0.439       \\ \hline
    111       &  0.061        &  0.047         &  0.206        \\ \hline
    134       &  0.033        &  0.035         &  0.138        \\ \hline
    161       &  0.063        &  0.063         &  0.388      \\ \hline
\end{tabular}
\end{center}
\label{individual_fitting_at_7tev_pp_collision_chi2_NDF}
\end{table}

\begin{table}[ht]
\caption{$\chi^{2}/\rm{NDF}$ of the strange hadrons production in pPb collision at 
$\sqrt{s_{NN}}$ = 5.02 TeV obtained using individual fitting}
\begin{center}
\begin{tabular}{| c || c | c | c |} 
\hline
&\multicolumn{3}{l|}{~~~~~~$\chi^{2}/\rm{NDF}$} \\ \hline 
$<N_{\rm{\rm trk}}>$ &  $K^{0}_{s}$    &  $\Lambda$     &  $\Xi^{-}$  \\     \hline\hline 
    21        &  0.043        &  0.029         &  0.058          \\ \hline 
    57        &  0.019        &  0.023         &  0.058        \\ \hline
    89        &  0.021        &  0.015         &  0.044       \\ \hline
    125       &  0.023        &  0.020         &  0.111        \\ \hline
    159       &  0.015        &  0.026         &  0.067        \\ \hline
    195       &  0.016        &  0.019         &  0.043        \\ \hline
    236       &  0.017        &  0.021         &  0.048        \\ \hline
    280       &  0.017        &  0.025         &  0.077        \\ \hline
\end{tabular}
\end{center}
\label{individual_fitting_at_502tev_pPb_collision_chi2_NDF}
\end{table}

\begin{table}[ht]
\caption{$\chi^{2}/\rm{NDF}$ of the strange hadrons production in PbPb collision at 
$\sqrt{s_{NN}}$ = 2.76 TeV obtained using individual fitting}
\begin{center}
\begin{tabular}{| c || c | c | c |} 
\hline
&\multicolumn{3}{l|}{~~~~~~$\chi^{2}/\rm{NDF}$} \\ \hline 
$<N_{\rm{\rm trk}}>$ &  $K^{0}_{s}$    &  $\Lambda$     &  $\Xi^{-}$  \\    \hline\hline 
    21        &  0.034        &  0.109         &  0.407           \\ \hline 
    58        &  0.024        &  0.069         &  0.507        \\ \hline
    92        &  0.033        &  0.077         &  0.387       \\ \hline
    130       &  0.068        &  0.029         &  0.204        \\ \hline
    168       &  0.023        &  0.185         &  0.249        \\ \hline
    210       &  0.063        &  0.028         &  0.285        \\ \hline
    253       &  0.068        &  0.056         &  0.336        \\ \hline
    299       &  0.080        &  0.059         &  0.179        \\ \hline
\end{tabular}
\end{center}
\label{individual_fitting_at_276tev_PbPb_collision_chi2_NDF}
\end{table}

\begin{table}[ht]
\caption{$\chi^{2}/\rm{NDF}$ of the strange hadrons production in pp collision at 
$\sqrt{s}$ = 7 TeV, pPb collision at $\sqrt{s_{NN}}$ = 5.02 TeV and PbPb collisions
at $\sqrt{s_{NN}}$ = 2.76 TeV obtained using combined fitting}
\begin{tabular}{| c | c | c | c | c | c |} 
\hline
\multicolumn{2}{|c|}{pp } & \multicolumn{2}{|c|}{pPb } & \multicolumn{2}{|c|}{PbPb } \\ \hline
$<N_{\rm{\rm trk}}>$& $ \frac{\chi^{2}}{\rm{NDF}}$&$<N_{\rm{\rm trk}}>$&$ \frac{\chi^{2}}{\rm{NDF}}$ 
&$<N_{\rm{\rm trk}}>$& $\frac{\chi^{2}}{\rm{NDF}}$ \\  \hline\hline
   14   &  1.989        &    21          &   1.035       &   21           &  1.954       \\ \hline 
   51   &  0.308        &    57          &   1.559       &   58           &  0.847       \\ \hline
   79   &  0.261        &    89          &   0.202       &   92           &  0.668      \\ \hline
   111  &  0.262        &    125         &   0.223       &   130          &  0.504      \\ \hline
   134  &  0.385        &    159         &   0.348       &   168          &  0.464      \\ \hline
   161  &  0.362        &    195         &   0.454       &   210          &  0.446      \\ \hline
    -   &       -       &    236         &   0.688       &   253          &  0.421      \\ \hline 
    -   &       -       &    280         &   0.833       &   299          &  0.509      \\ \hline
\end{tabular}
\label{strange_particle_production_PbPb_collision_chi2_NDF}
\end{table}

\section{Conclusion}

We carried out an analysis of transverse momentum spectra of the strange hadrons
in different multiplicity events produced in pp collision at $\sqrt{s}$ = 7 TeV, 
pPb collision at $\sqrt{s_{NN}}$ = 5.02 TeV and PbPb collision at $\sqrt{s_{NN}}$ = 
2.76 TeV using Tsallis distribution modified to include transverse flow. 
The analysis is performed for both the differential and single freeze out scenarios 
of the strange hadrons $K^{0}_{s}, \Lambda$ and $\Xi^{-}$. 
In both freeze-out scenarios, the transverse flow increases with event multiplicity 
for all the three systems.

In the case of the differential freeze-out scenario, the value of parameter $n$ 
has a little variation with multiplicity implying that the degree of thermalization 
remains similar for the events of different multiplicity classes in all the three 
systems. 
For the differential freeze-out scenario the value of Tsallis temperature increases 
with the mass of the hadrons and also increases with the multiplicity of the event 
class in pp and pPb system. 
In PbPb system, the value of Tsallis temperature  shows little variation with the 
multiplicity for all the three hadrons.

In the case of the single freeze-out scenario the difference between small systems 
(pp, pPb) and PbPb system becomes more evident. 
For pp and pPb systems, the parameter $n$ decreases with increasing multiplicity 
but for PbPb system it shows a slight increase with the multiplicity and has a 
larger value as compared to that for smaller systems.  
For single freeze-out scenario, the temperature in PbPb system smoothly increases 
with multiplicity for the first few bins then smoothly decreases which is a trend 
opposite to what is found in the pp and pPb systems. 
It is expected that if a truly collective system is formed, the temperature
decreases or remains constant as the system size increases. 
Decreasing temperature for PbPb system above $N_{\rm trk}$=150 is indicative of that.

\section*{Acknowledgement}

We are thankful to CMS publication committee for providing the tables of the 
published data.

\end{document}